\documentclass[traditabstract]{aa}

\usepackage[english]{babel}
\usepackage[ansinew]{inputenc}
\usepackage{amssymb,amsmath}
\usepackage{graphicx} 
\usepackage{natbib} 

\title{Assessment of stochastic and deterministic models of 6\,304~quasar lightcurves from SDSS Stripe~82}
\titlerunning{Assessment of stochastic and deterministic models of QSO lightcurves}

\author{Ren\'e Andrae\inst{1}\thanks{andrae@mpia-hd.mpg.de} \and Dae-Won Kim\inst{1} \and Coryn A.L.~Bailer-Jones\inst{1}}
\authorrunning{R.~Andrae et al.~(2013)}
\date{Accepted by A\&A on Apr 04 2013}
\institute{${}^1$ Max-Planck-Institut f\"ur Astronomie, K\"onigstuhl 17, 69117 Heidelberg, Germany}

\abstract{The optical light curves of many quasars show variations of tenths of a magnitude or more on time scales of months to years. This variation often cannot be described well by a simple deterministic model. We perform a Bayesian comparison of over 20 deterministic and stochastic models on 6\,304~QSO light curves in SDSS Stripe 82. We include the damped random walk (or Ornstein--Uhlenbeck [OU] process), a particular type of stochastic model which recent studies have focused on. Further models we consider are single and double sinusoids, multiple OU processes, higher order continuous autoregressive processes, and composite models. We find that only 29~out of 6\,304~QSO lightcurves are described significantly better by a deterministic model than a stochastic one. The OU process is an adequate description of the vast majority of cases (6\,023). Indeed, the OU process is the best single model for 3\,462 light curves, with the composite OU process/sinusoid model being the best in 1\,706 cases. The latter model is the dominant one for brighter/bluer QSOs. Furthermore, a non-negligible fraction of QSO lightcurves show evidence that not only the mean is stochastic but the variance is stochastic, too. Our results confirm earlier work that QSO light curves can be described with a stochastic model, but place this on a firmer footing, and further show that the OU process is preferred over several other stochastic and deterministic models. Of course, there may well exist yet better (deterministic or stochastic) models which have not been considered here.}


\begin{document}
\maketitle
%


\section{Introduction}
\label{sect:introducation}

The study of quasar lightcurves has prospered in the last years with the advent of multi-epoch observations such as SDSS Stripe~82 or MACHO. \citet{Sesar2007} have shown that more than 90\% of QSOs in SDSS Stripe~82 exhibit time variations of 0.03~mag in the optical on timescales of a few years. This time variation is one of the most important characteristics of QSOs and is widely believed to be associated with accreting material falling into the central black holes of QSOs \citep{Rees1984, Kawaguchi1998}.  Many authors \citep{Kozlowski2010, MacLeod2010, Schmidt2010, Butler2011, Kim2011, Kim2012, Pichara2012} have shown that QSO variability can be used to discriminate QSOs from other sources. Some of these works also showed possible correlations between the time variation and physical characteristics of QSOs, such as black hole mass, using a stochastic model \citep{Kelly2009}. This suggests that modelling QSO lightcurves will help us to understand the cause of QSO variability. Thus analysing QSO lightcurves is critical not only for efficient QSO selection but also for a better understanding of the QSO variability mechanism.

In this article, we model quasar lightcurves that have been observed in SDSS Stripe~82 \citep{Ivezic2007}. Previous investigations \citep[primarily][]{Kelly2009} have successfully employed a ``damped random walk'' as a stochastic model for QSO lightcurves from MACHO data. This model was motivated by earlier results that showed that QSO lightcurves typically exhibit power spectra that are power laws with slopes $\approx 2$ \citep{Giveon1999,Collier2001}, a behaviour that can be achieved with a damped random walk in certain domains, too. \citet{Kozlowski2010} and \citet{MacLeod2010} confirmed that the damped-random-walk model of \citet{Kelly2009} indeed provides a good description of QSO lightcurves. \citet{Zu2013} performed a model comparison of power spectra of 55 OGLE quasars but could not detect any significant deviations from the damped random walk using frequentist hypothesis testing. However, no rigorous Bayesian model comparison based on the light curves' photometry itself has been undertaken to date.

The principal objective of this article is to conduct a Bayesian model assessment on a set of 6\,304~QSO lightcurves of \citet{Ivezic2007}. We try to assess whether the damped random walk is indeed a reliable model for QSO lightcurves, in comparison to numerous other models. \citet{Kelly2009} speculated whether extensions of this process may lead to even better models. Therefore, we also compare the damped-random-walk model to several of its extensions. On a more fundamental level, we want to assess whether QSO variability is deterministic or stochastic.

We present various deterministic and stochastic models for QSO lightcurves in Sect.~\ref{sect:models}. We discuss the data and its properties in Sect.~\ref{sect:data} and present our results in Sect.~\ref{sect:results}. Finally, we conclude in Sect.~\ref{sect:conclusions}.

\section{Models and model comparison}
\label{sect:models}

In this section, we discuss the method for model comparison and briefly summarise the models we are considering.

\subsection{Bayes factors}
\label{sect:intro-Bayes-factors}

We perform the model comparison by means of Bayes factors. We explicitly decide against using $p$-values in orthodox hypothesis tests (reduced $\chi^2$/$\chi^2$-test, KS-test, $F$-test, etc.) because they are known to favour increasingly complex models as the amount of observational data increases \citep[e.g.][]{Berger1987,Kass1995,Berger2003,Christensen2005}. In particular, as more data becomes available, $p$-values generally do \textit{not} converge to favour the ``true'' model. Therefore, if a model is rejected by a $p$-value, we can never be certain whether the model is indeed inappropriate, or whether it is this inherent failure of the $p$-value itself at work. Bayes factors do not suffer from this problem \citep[e.g.][]{Kass1995}.

Let $D$ denote the observed data and let $M_0$ and $M_1$ denote two models with model parameters $\theta_0$ and $\theta_1$, respectively. The Bayes factor is then defined as the ratio of Bayesian evidences,
\begin{equation}\label{eq:def:Bayes-factor}
\frac{p(D|M_1)}{p(D|M_0)} = \frac{\int d\theta_1\,p(D|\theta_1,M_1)\,p(\theta_1|M_1)}{\int d\theta_0\,p(D|\theta_0,M_0)\,p(\theta_0|M_0)} \,\textrm{,}
\end{equation}
where $p(\theta|M)$ denotes the prior distribution of the model parameters $\theta$ of model $M$. In words, the Bayesian evidence $p(D|M)$ of the data $D$ given a model $M$ is the probability that the observed data $D$ could have been generated from model $M$, \textit{irrespective} of the parameter values of model $M$.

Bayes factors penalise model complexity in a very intuitive way. As the Bayesian evidence is a prior-weighted mean of the likelihood function, an additional model parameter has to lead to an increase in average likelihood. Therefore, an additional model parameter has to provide a ``net increase'' in likelihood in order to be ``plausible''.

Estimation of the Bayes factor is nontrivial due to the marginalisation integrals of the Bayesian evidences in Eq.~(\ref{eq:def:Bayes-factor}). Therefore, we estimate the Bayesian evidence by Monte Carlo integration: We draw random samples from the prior distribution, evaluate the likelihood function for the sampled parameter values, and compute the mean value of all likelihoods.

\subsection{Interpretation of Bayes factors}
\label{sect:interpretation-log-BF}

Given two models, $M_0$ and $M_1$, and estimates of their Bayesian evidences, $p(D|M_0)$ and $p(D|M_1)$, we use the logarithmic Bayes factor,
\begin{equation}
\Delta_{1-0}=\log_{10}p(D|M_1)-\log_{10}p(D|M_0) \,\textrm{.} \\
\end{equation}
Throughout this work, we use the terminology of \citet{Kass1995}, where $\Delta_{1-0} > 2$ is interpreted as ``decisive evidence against $M_0$''. In words, this means the observed data is a factor $>100$ more likely to have been generated by model $M_1$ than by model $M_0$, irrespective of those models' parameter values.

A (logarithmic) Bayes factor of $\Delta_{1-0} > 2$ can only be interpreted as decisive evidence against $M_0$, since we have found at least one other model (namely $M_1$) that outperforms $M_0$. However, we cannot conclude from $\Delta_{1-0} > 2$ that we have found decisive evidence in favour of $M_1$. There might be other models that we have not tested that would outperform $M_1$. In order to provide evidence \textit{in favour} of some model, we would have to test \textit{all} reasonable alternatives. In practice, the set of reasonable alternatives is infinitely large and cannot be covered.

\subsection{Prior distributions and Bayes factors}

The choice of prior distributions on model parameters obviously has a large impact on the Bayesian evidence and any Bayes factor computed from it. Prior distributions therefore need to be chosen carefully.\footnote{\citet{BailerJones2012} introduces a formalism for Bayesian model comparison based on K-fold cross validation that is less sensitive to the choice of prior distributions. However, we decided not to use this approach because it would be numerically too expensive for our application.} As we have no intuition for most of the models we are considering, we decided to randomly select 100 QSO lightcurves from our data sample and to perform a maximum-likelihood parameter estimation. The resulting distribution of best-fit parameters over this subset is then fitted by an appropriate model, e.g., a Gaussian, a log-normal, or a uniform distribution. These distributions are then employed as prior distributions on all QSO lightcurves. This procedure provides us with a coherent choice of prior distributions, based on the characteristics of the ensemble of lightcurves. We emphasise that we do \textit{not} use the parameter values reported by \citet{Kelly2009} to infer prior distributions for the damped-random-walk model.

\subsection{Deterministic vs.~stochastic models}

We broadly distinguish between ``deterministic'' and ``stochastic'' models. For a deterministic model, e.g., a sinusoid, we can predict the model's (not the data's) time evolution without uncertainty. Conversely, for a stochastic model, e.g., a random walk, this prediction is only probabilistic, even when we know the true model parameters.

\subsection{Deterministic models}

As alternatives to stochastic processes in general, we are considering a few simple deterministic models for QSO lightcurves.

\subsubsection{Constant model}

The simplest possible model of a QSO lightcurve $m(t)$ is a constant model,
\begin{equation}
m(t) = a_0 \;\textrm{.}
\end{equation}
This model assumes that there is no intrinsic time variability in the QSO lightcurve itself and any fluctuation is attributed to the measurement error. Given the results from \citet{Sesar2007}, it is obvious that this model may not be a very useful description of QSO lightcurves. However, it is the simplest model and should thus be taken into consideration.

\subsubsection{Constant model plus additional noise}

This model is an extension of the constant model that adds an additional noise term, which is independent of the measurement error. This additional free parameter can be interpreted either as intrinsic to the QSO lightcurve itself or as correction for a potentially underestimated measurement error, or both. We have no possibility of differentiating between these two interpretations, without having additional information.

\subsubsection{Sinusoidal models}

The sinusoidal model assumes that the QSO lightcurve is periodic around a mean value, $a_0$,
\begin{equation}
m(t) = a_0 + a_1\cos(\omega\,t+\phi) \;\textrm{.}
\end{equation}
We use this model with flat priors in period and in logarithmic period for periods between 1~and 10\,000~days, i.e., this model appears twice. Furthermore, we consider the sinusoid with a periodogram prior. Note that by using a frequentist periodogram \citep{Horne1986} as prior distribution, we already make excessive use of the available data. Therefore, this prior distribution is highly favourable for this model.\footnote{That argument assumes that a periodogram favours periods that also achieve high likelihoods for the given data.} This is acceptable because we are going to find this model to have low evidence later, such that this prior in favour of this model is a conservative choice. We also consider the model comprised of a double sinusoid, where the prior is uniform over period for both components.

\subsubsection{Sinusoidal model plus additional noise}

This model is an extension of the single-sinusoid model that adds an additional noise term, which is independent of the measurement error. Again, this additional free parameter can be interpreted either as intrinsic to the QSO lightcurve itself or as correction for a potentially underestimated measurement error, or both.

\subsection{Stochastic models}

We consider various different stochastic models. Evidently, we cannot lay out the details of all these processes here. Instead, the interested reader may refer to \citet{Brockwell2002} for more details.

\subsubsection{Damped-random-walk model (OU process)}

The chief character in this work is the ``damped random walk''. This model is actually called the Ornstein-Uhlenbeck process \citep[][hereafter OU process]{Uhlenbeck1930,Gillespie1996} and we adopt this name for the rest of this work. The OU process is the only stochastic process that is Markov, Gaussian and stationary. It is a special kind of continuous-time first-order autoregressive process, which is abbreviated as CAR(1). These are defined by the recursion formula
\begin{equation}\label{eq:recursion_relation_CAR1}
Y_n = \phi(t_n,t_{n-1})Y_{n-1} + Z_n \,\textrm{,}
\end{equation}
where $Y_n$ is the observed value at time $t_n$, $\phi(t_n,t_{n-1})$ is a deterministic function, and $Z$ is a random variate ``driving'' the stochastic process. Such a process is called ``autoregressive'', because $Y_n$ can be predicted from the previous observation $Y_{n-1}$. It is first order, because $Y_n$ is predicted from $Y_{n-1}$ only, while $Y_{n-2}$ etc.~have no influence once $Y_{n-1}$ is given. It is continuous time, because the observation times $t_n$ are non-uniformly spaced. For the OU process, $Z$ is a Gaussian random variate and a special choice for $\phi$ is made,
\begin{equation}\label{eq:Kellys-relaxation-timescale}
\phi_\textrm{OU}(t_n,t_{n-1}) = \exp\left[-\frac{|t_n - t_{n-1}|}{\tau}\right] \,\textrm{,}
\end{equation}
where $\tau$ is the relaxation timescale. More mathematical details on the OU process, its likelihood function and parameter estimation, can be found, e.g., in \citet{Kelly2009} and, in more detail, in \citet{BailerJones2012}. We compute likelihoods of the OU process as described in Sect.~3.1 of \citet{Kelly2009}.

\subsubsection{Extending the OU process}

\begin{figure}
\includegraphics[width=8.0cm]{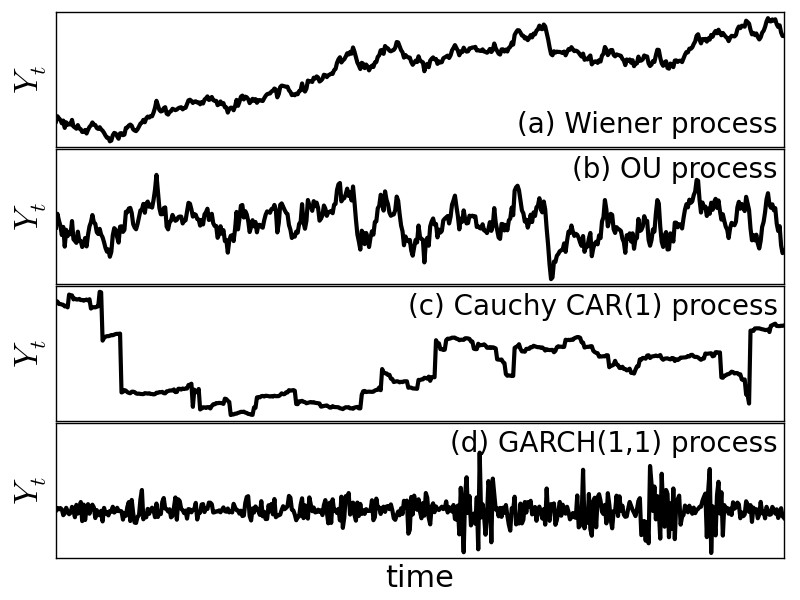}
\caption{Examples of stochastic processes. We can see qualitative differences by eye: The Wiener process in panel (a) drifts away, whereas the OU process in panel (b) reverts to the mean. The Cauchy process in panel (c) has ``steps'' due to the heavy tails of the Cauchy distribution. The GARCH(1,1) process in panel (d) exhibits episodes of low and high variation.}
\label{fig:examples-stochastic-processes}
\end{figure}

\citet{Kelly2009} speculate that extensions of the OU process could provide improved descriptions of QSO lightcurves. However, there are various different ``dimensions'' along which we can extend the OU process. Since the OU process is a Gaussian CAR(1) process with a special choice of $\phi$, possible extensions are:
\begin{itemize}
\item Wiener process, which is the Gaussian CAR(1) process with $\phi=1$. This is the classical random walk.
\item Linear combinations of multiple OU processes. Likelihoods are evaluated using the Kalman recursion as described in \citet{Kelly2011}. In particular, the component OU processes have identical driving amplitudes and vary only in their individual timescales, as in \citet{Kelly2011}.\footnote{In fact, each OU process could also have its own amplitude but this would make the model even more complex.}
\item Gaussian CAR(2) processes with two free deterministic functions, $\phi_1(t_i,t_j)$ and $\phi_2(t_i,t_j)$ for which we consider three different parameterisations. This is a second-order extension of the OU process, in which Eq.~(\ref{eq:recursion_relation_CAR1}) is extended by another predecessor. First, we choose constant $\phi_1=\phi_2=0.1$, which we found to yield relative high likelihood values. Second, we choose a parametrisation as in Eq.~(\ref{eq:Kellys-relaxation-timescale}), where the two relaxation timescales $\tau_1\neq\tau_2$ are allowed to be different. Third, we choose two relaxation timescales that are identical, i.e., $\tau_1=\tau_2$.
\item Gaussian CARMA(1,1) process, which extends the OU process by a first-order moving average. This introduces a smoothing of random fluctuations and thus corresponds to another kind of ``damping''.
\item Non-Gaussian CAR(1) processes with $\phi=1$ (``non-Gaussian random walk''). We consider only $\alpha$-stable distributions, namely the Cauchy distribution, the symmetric stable distribution, and the (generally asymmetric) stable distribution. Appendix~\ref{app:stable-PDFs} provides a brief overview of stable distributions.
\item Gaussian ARCH(1) \citep{Engle1982} and GARCH(1,1) processes \citep{Bollerslev1986}, where there is stochasticity in the variance instead of the mean. These stochastic processes are also combined with aforementioned processes, providing models such as CARMA-GARCH with stochasticity in both mean and variance.
\end{itemize}
Figure~\ref{fig:examples-stochastic-processes} shows examples of some of these stochastic processes.

\section{The data}
\label{sect:data}

\subsection{QSO lightcurves from SDSS Stripe~82}

\begin{figure}
\includegraphics[width=8.0cm]{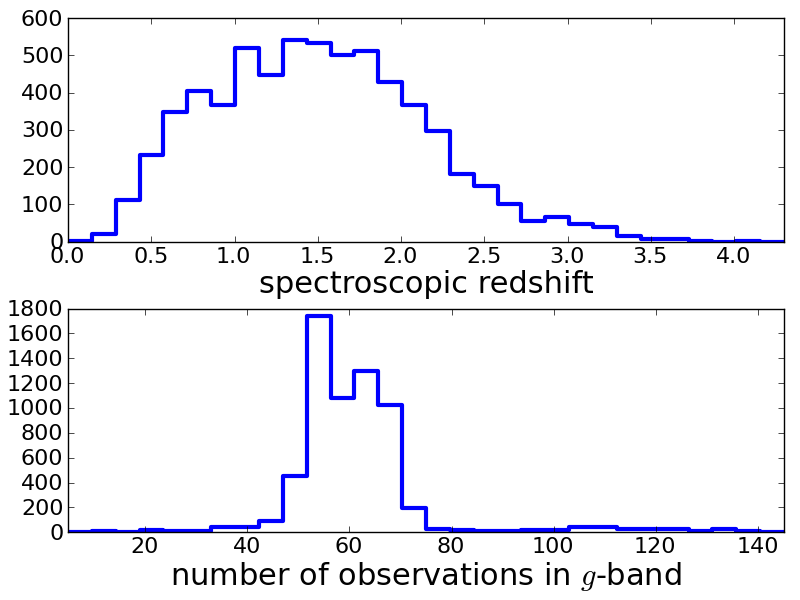}
\caption{Sample distributions of spectroscopic redshift estimates (top panel) and number of observations in the $g$-band lightcurve (bottom panel).}
\label{fig:sample-distribution-z-Nobs}
\end{figure}

\citet{Ivezic2007} compiled a database of lightcurves of variable objects from the SDSS Stripe~82 data. From these data, we extract quasars using the spectroscopic redshift estimate. Any variable object with redshift $z\geq 0.08$ is considered as a quasar. This provides us with a sample of 6\,304~QSOs, the redshift distribution of which is shown in the top panel of Fig.~\ref{fig:sample-distribution-z-Nobs}. From these data, we select only the $g$-band lightcurves, since this is one of the deepest bands and has reliable magnitude estimates. We do \textit{not} conduct a multivariate time-series analysis, first, since such models are considerably more complex and thus harder to constrain from the available data, and second, since the different bands exhibit strongly correlated variability \citep{Schmidt2012}. Consequently, studying multi-band variability is unlikely to provide considerably more insight than a single-band analysis. The bottom panel of Fig.~\ref{fig:sample-distribution-z-Nobs} shows that the majority of $g$-band lightcurves has between 40 and 80 observations. The minimum number of observations is 11 and the maximum number is 140.

\subsection{Accounting for measurement errors}

We neglect measurement errors on the time domain $t_n$ because they are very small. However, we cannot neglect measurement errors on the observed magnitudes $g_n$. We assume that these errors are Gaussian standard deviations, $\sigma_n$.

Accounting for such measurement errors for deterministic models leads to a likelihood function. However, in the case of stochastic models, data and model are \textit{both} probability distributions. Consequently, the likelihood of a single observed magnitude $g_n$ is given by a convolution of model and error distribution,
\begin{equation}\label{eq:model-convolution}
\mathcal L(g_n|\sigma_n,\theta) = \int_x \mathcal N(x|\mu=g_n,\sigma_n)\,p(x|\theta)\,dx \,\textrm{,}
\end{equation}
where $\mathcal N$ denotes the normal distribution and $p(x|\theta)$ is the stochastic model of $x$ with parameter values $\theta$ \citep[e.g., see][]{BailerJones2012}.

If we consider Gaussian processes, the convolution integral of Eq.~(\ref{eq:model-convolution}) is analytic. However, in the case of a stable process -- Cauchy, symmetric, or general -- this convolution is not analytic. In our implementation, the integral is then estimated by drawing 100 Monte Carlo samples from the Gaussian error distribution and evaluating the likelihood of the stable process.

\subsection{Outlier measurements}

Outliers in the measured lightcurves pose a severe problem to any analysis. As stable distributions have heavier tails than the Gaussian distribution, we could not tell a-priori whether such outliers were measurement flukes or real observations. Handling such outliers would be nontrivial because excluding them from the data analysis would possibly deprive us of our most valuable data if QSO lightcurves were stable processes. It would be necessary to add an outlier distribution to the Gaussian stochastic process, though it is unclear how such an outlier distribution should be constructed. Therefore, we only conduct a coarse outlier removal on the data, excluding any measurement with extreme $g$ measurement or extreme measurement error. We consider all remaining data from SDSS Stripe~82 as valid measurements, i.e., we take the remaining data at face value and assume that there are no outliers. Consequently, if we find the Cauchy process not to be a good description despite of taking all data at face value, this will be a conservative result.

\section{Results}
\label{sect:results}

\subsection{Decisive evidence with confidence levels}

We normally consider a model $M_1$ to provide decisive evidence against model $M_0$, if the logarithmic Bayes factor $\Delta_{1-0}$ is larger than 2 (Sect.~\ref{sect:interpretation-log-BF}). However, by estimating the Bayesian evidence through Monte Carlo integration, we introduce a random error into the estimate of $\Delta_{1-0}$, which we denote by $\sigma_\Delta$. If $\sigma_0$ and $\sigma_1$ denote the statistical errors of the Monte Carlo estimates of Bayesian evidences, we estimate $\sigma_\Delta$  through error propagation,
\begin{equation}
\sigma_\Delta^2 = \sigma_0^2 + \sigma_1^2 \,\textrm{.}
\end{equation}
Typically, $\sigma_\Delta$ is of the order of 0.5 or less, such that the criterion $\Delta>2$ may not be as confident as we expect. In order to account for the uncertainty in our estimate of the Bayes factor, we modify our criterion: We consider a model $M_1$ to provide ``decisive evidence against $M_0$ with $3\sigma$ confidence'', if
\begin{equation}
\Delta_{1-0} > 2+3\sigma_\Delta \,\textrm{.}
\end{equation}
We likewise define decisive evidence with $5\sigma$, $7\sigma$, and $10\sigma$ confidence.

\begin{table*}
\begin{centering}
\begin{tabular}{lcrrrrr}
\hline
model & 
\rotatebox{80}{number of model parameters} & 
\rotatebox{80}{no decisive evidence against it at $3\sigma$} &
\rotatebox{80}{no decisive evidence against it at $5\sigma$} & 
\rotatebox{80}{no decisive evidence against it at $7\sigma$} & 
\rotatebox{80}{no decisive evidence against it at $10\sigma$} & 
\rotatebox{80}{max evidence}\\
\hline\hline
constant                                                  & 1 & 0 & 0 & 0 & 0 & 0 \\
constant plus noise                               & 2 & 59 & 64 & 73 & 94 & 4 \\
sinusoid (flat prior in $P$)                              & 4 & 28 & 37 & 46 & 58 & 2 \\
sinusoid (flat prior in $\log P$)                         & 4 & 20 & 23 & 29 & 39 & 2 \\
sinusoid (periodogram prior)                              & 4 & 27 & 30 & 39 & 50 & 4 \\
two sinusoids (flat priors in $P$)                        & 7 & 33 & 36 & 45 & 60 & 7 \\
sinusoid (flat prior in $P$) plus noise                              & 5 & 467 & 583 & 721 & 921 & 149 \\
\hline
Wiener process (random walk)                     & 2 & 2205 & 2724 & 3230 & 3949 & 11 \\
OU process (damped random walk)                     & 3 & 6023 & 6069 & 6093 & 6131 & 5047 \\
Gaussian CAR(2) process ($\phi_1=\phi_2=0.1$)             & 2 & 0 & 0 & 0 & 1 & 0 \\
Gaussian CAR(2) process ($\tau_1=\tau_2$)                 & 3 & 1 & 2 & 2 & 3 & 0 \\
Gaussian CAR(2) process ($\tau_1\neq\tau_2$)              & 4 & 5 & 5 & 5 & 5 & 5 \\
Cauchy CAR(1) process ($\phi=1$)                          & 2 & 81 & 108 & 172 & 361 & 36 \\
symmetric stable CAR(1) process ($\phi=1$, $\beta=0$)     & 3 & 0 & 0 & 0 & 0 & 0 \\
stable CAR(1) process ($\phi=1$)                          & 4 & 0 & 0 & 0 & 0 & 0 \\
Gaussian CARMA(1,1) process                       & 4 & 0 & 0 & 0 & 0 & 0 \\
\hline
constant + Gaussian ARCH(1) process                     & 3 & 236 & 268 & 306 & 355 & 69 \\
sinusoid + Gaussian ARCH(1) process                     & 6 & 923 & 1208 & 1529 & 2096 & 155 \\
Gaussian CAR(1)-ARCH(1) process                     & 4 & 1884 & 2257 & 2639 & 3183 & 219 \\
Gaussian CAR(1)-GARCH(1,1) process                     & 5 & 3542 & 3912 & 4285 & 4751 & 536 \\
Gaussian CARMA(1,1)-GARCH(1,1) process           & 6 & 1533 & 2041 & 2593 & 3486 & 58 \\
\hline
\end{tabular}
\end{centering}
\caption{Results of evidence-based model comparison. For each model under consideration we quote the number of model parameters, the number of QSO lightcurves where there is no decisive evidence against this model at a specified confidence level, and the number of QSO lightcurves where each model provides the highest evidence. The upper block of the table contains deterministic models. The middle block contains stochastic models with constant variance, while the lower block contains stochastic process with stochastic variance. (Periods of all sinusoids are between 1 and 10\,000 days.) The last column adds up to the total of 6\,304~QSO lightcurves, the other columns do not.}
\label{table:results-model-comparison}
\end{table*}

\subsection{Overview}

Table~\ref{table:results-model-comparison} summarises the results of comparing the models we are considering. Inspecting the last column of maximum evidences, two things are immediately obvious: First, the OU process is clearly superior to all other models considered here, providing the best model for 5\,047~out of 6\,304~QSO lightcurves. Second, the deterministic models considered here do a rather poor job in describing those QSO~lightcurves.

Let us now study Table~\ref{table:results-model-comparison} in more detail. First, we consider the deterministic models. For \textit{all} QSO lightcurves, there is decisive evidence against the model of a constant lightcurve at $10\sigma$ confidence. Adding an additional noise term to the constant model improves the results considerably but still provides a good description only for a handful of QSO lightcurves. Sinusoidal models are similarly poor descriptions. In particular, we observe that the periodogram prior does not perform any better than the other priors on period. Given the limits on period (1 to 10\,000 days), the flat prior in $P$ seems to outperform the flat prior in $\log P$, which may indicate that if QSO lightcurves were single sinusoids, they would prefer large periods, since the flat prior in $P$ favours large values of $P$ compare to a prior flat in $\log P$. Adding yet another sinusoidal component also does not improve the model performance considerably.

Now let us consider Table~\ref{table:results-model-comparison} with respect to the stochastic models. For a significant fraction of QSO lightcurves, there is no decisive evidence against the Wiener process with confidence, though only few QSOs are best fit by this model. Nevertheless, the OU process clearly outperforms the Wiener process. This is astrophysically sensible because the (marginal) variance diverges with time for the Wiener process, yet the QSO only has finite energy supply. For 6\,023 lightcurves, no model can provide decisive evidence against the OU process at any relevant confidence level. Next, Gaussian CAR(2) processes apparently do not improve the results obtained by the OU process. Stable CAR(1) processes  generally seem to provide very poor descriptions, though the Cauchy CAR(1) process is favoured by some lightcurves which may indicate the presence of outlier measurements in these lightcurves. However, there is decisive evidence against general stable CAR(1) processes for all lightcurves. Similarly, the CARMA(1,1) process is outperformed by the OU process. Last but not least, we note that adding stochastic variance (ARCH and GARCH) does improve the modelling outcome. In particular, extending the OU process by a GARCH(1,1) component, we obtain a model that is almost as good as the OU process alone. This suggests that some QSOs exhibit stochasticity also on the variance not only on the mean.

Given the infinite (marginal) variance of the Wiener process and the heavy tails of the Cauchy CAR(1) process, respectively, we tested whether the preference for any of these two models shows any dependence on outlier measurements. In particular, we find that preference for these two models does not appear to be connected with maxmium $g$-band measurement errors or maximum absolute measurement-error-weighted difference to the median magnitude of the lightcurve. However, we do find a clear tendency that preference for the Wiener process or the Cauchy CAR(1) process is more likely for QSO lightcurves with fewer observations and at lower redshfits.

As an additional sanity check, we repeated the analysis of the OU process, in which the prior distributions were twice and half as broad as before. In both cases, we find no noteworthy changes in Table~\ref{table:results-model-comparison}. This suggests that the choice of prior distributions only has a minor impact on the preference for the OU process.

\begin{figure*}
\begin{center}
\includegraphics[width=16cm]{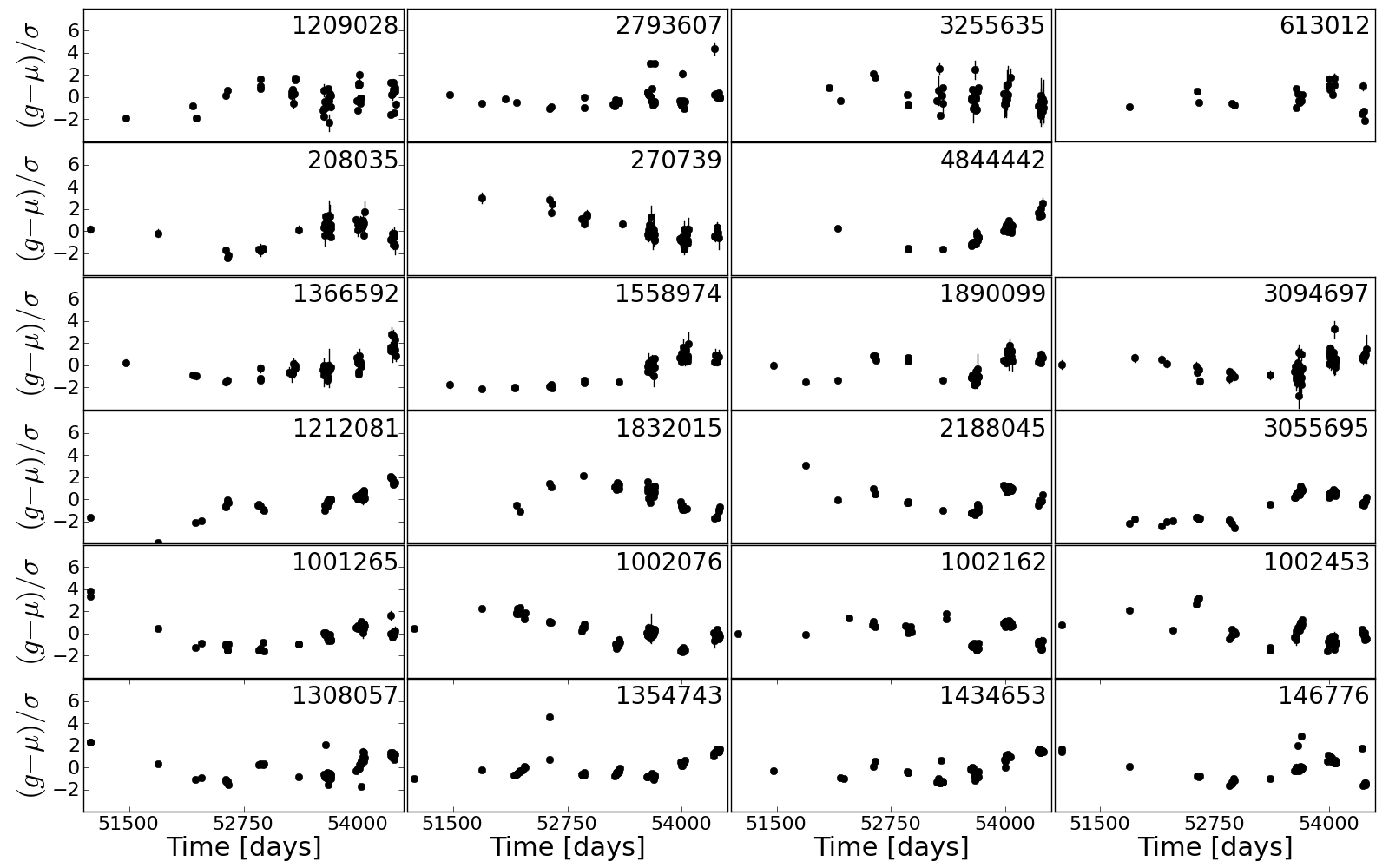}
\end{center}
\caption{Example lightcurves with maximum evidence for constant model with additional noise (first row), for sinusoid with flat prior in period (second row), for two sinusoids (third row), for Wiener process (fourth row), for OU process (fifth row), and for Cauchy CAR(1) process (sixth row). In order to facilitate comparison for this figure, all lightcurves have been standardised by subtracting the mean and dividing by the standard devition. Numbers in the top right corners of each panel provide this QSO's ID from \citet{Ivezic2007}.}
\label{fig:examples-lightcurves}
\end{figure*}

Figure~\ref{fig:examples-lightcurves} shows example lightcurves of four QSOs that favour certain models. We note that lightcurves in the top three rows exhibit strong variation in magnitude over short timescales. Lightcurves in the bottom panels also may exhibit large variations in magnitude, but over longer timescales. There are no obvious periodicities in any of these lightcurves. Lightcurves in the last row exhibit a few outlier observations, which may explain why they favour a Cauchy CAR(1) process with heavy tails.

\subsection{Assessing stochasticity}

Table~\ref{table:results-model-comparison} clearly shows that the OU process is by far the best model for QSO lightcurves of those models considered so far. We now want to generally assess the stochasticity of QSO lightcurves by comparing the combined deterministic vs.~the combined stochastic models. For this purpose, we compute the number of QSOs where there is at least one deterministic model where no stochastic model can provide decisive evidence against it, and vice-versa. Table~\ref{table:stochasticity-assessment} shows the results. Although there are 171 out of 6\,304~QSO lightcurves that are best described by a determinisitc model, there are only 29~QSO lightcurves where no stochastic models can provide decisive evidence against it at all relevant confidence levels. Conversely, there are 5\,353~QSO lightcurves where a stochastic model provides decisive evidence against all deterministic models at $10\sigma$ confidence. This provides a strong indication that QSO lightcurves are of stochastic nature.

\begin{table}
\begin{centering}
\begin{tabular}{ccc}
\hline
decisive evidence & deterministic & stochastic \\
against model & models & models  \\
\hline\hline
at $3\sigma$ & 5811 & 29\\
at $5\sigma$ & 5708 & 17\\
at $7\sigma$ & 5565 & 15\\
at $10\sigma$ & 5353 & 9\\
\hline
maximum evidence & 171 & 6133\\
\hline
\end{tabular}
\end{centering}
\caption{Assessment of deterministic vs.~stochastic models. For instances, stochastic models provide decisive evidence against all deterministic models at $10\sigma$ confidence level for 5\,353 QSO lightcurves.}
\label{table:stochasticity-assessment}
\end{table}

\subsection{Linear combinations of OU processes}

\citet{Kelly2011} suggest to use linear combinations of multiple OU processes. They find that typically a mixture of $\gtrsim 30$ OU processes is necessary to fit the observed power spectra of their QSO light curves. However, both methods for estimating the number of OU processes outlined in \citet{Kelly2011} (Sect.~4.2 therein) do not correctly penalise model complexity \citep[see Sect.~\ref{sect:intro-Bayes-factors} and discussions in][]{Berger1987,Kass1995,Berger2003,Christensen2005}. Consequently, using a linear combination of 30 OU processes or more is very likely to be an ``overfit''.\footnote{Models of arbitrary complexity can be correctly and consistently dealt with in a Bayesian model comparison. However, from an astrophysical point of view, such a complex model is a-priori rather implausible.}

Using the 6\,304~QSO lightcurves from SDSS Stripe~82, we want to assess how many OU processes can be superposed without overfitting the data.  Table~\ref{table:multiple-OU} shows that for the vast majority of QSO lightcurves, there is no decisive evidence against the single OU process. There are only very few lightcurves that favour two or more OU processes. Again, as a sanity check, we repeated the analysis with prior distributions half and twice as broad. We do not find any noteworthy change in our results, which implies that we are not overly sensitive to the choice of prior distributions. Unfortunately, our results are not directly comparable to \citet{Kelly2011}, since our lightcurves do not contain as many observations (c.f.~Fig.~\ref{fig:sample-distribution-z-Nobs}).

\begin{table}
\begin{centering}
\begin{tabular}{ccc}
\hline
number of    	& no decisive evidence 	& maximum \\
OU processes 	& against this model 	& evidence \\
                      	& at $3\sigma$ confidence 	&  \\
\hline\hline
1 & 6281 & 6274  \\
2 & 39 & 26  \\
3 & 39 & 3  \\
4 & 37 & 0  \\
5 & 38 & 0  \\
6 & 37 & 0  \\
7 & 37 & 0  \\
8 & 37 & 0  \\
9 & 37 & 1  \\
\hline
\end{tabular}
\end{centering}
\caption{Estimating the optimal number of linear combinations of OU processes.}
\label{table:multiple-OU}
\end{table}

\subsection{Parameters of OU process}

As we found the OU process to be by far the best model for QSO lightcurves, let us consider this model in more detail. We first show some example lightcurves with their best-fit OU process, and then investigate the distribution of best-fit parameter values.

So far, in order to evaluate Bayesian evidences, there was no need to actually maximise the likelihood function. However, now this becomes necessary. Starting out from the 20\,000 Monte Carlo samples drawn from the prior PDF, we select the parameter combination that achieved the highest likelihood. We then initialise a Simplex algorithm \citep{Nelder1965} at this parameter combination and let it further maximise the likelihood function.

\begin{figure}
\begin{center}
\includegraphics[width=8.0cm]{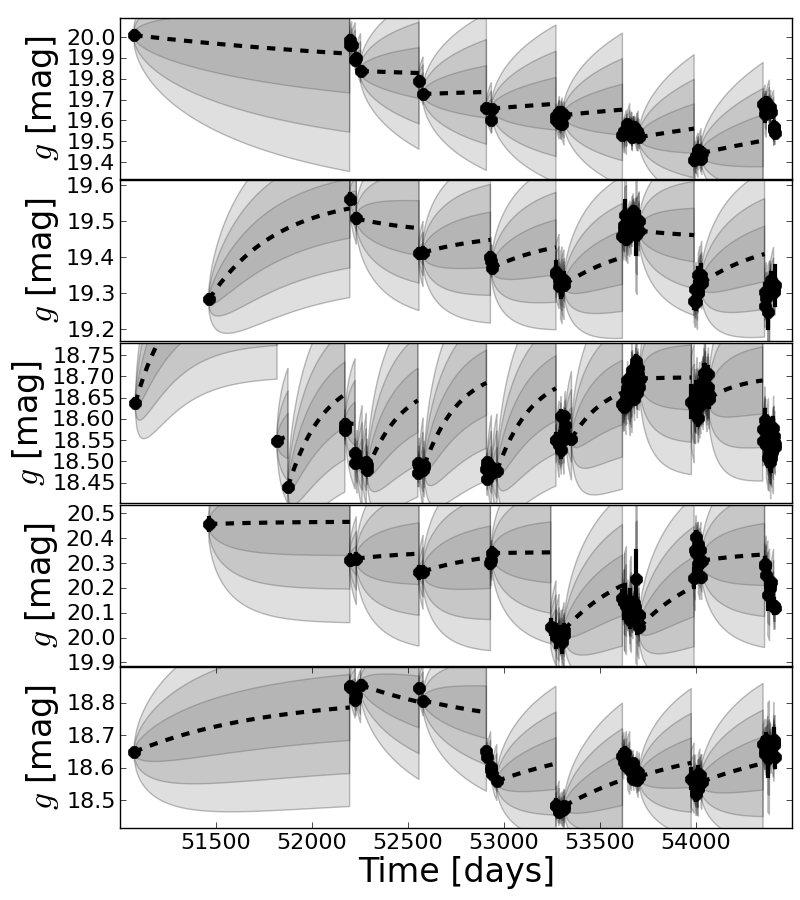}
\end{center}
\caption{Five randomly selected example lightcurves where the OU process has highest evidence of all models. The best-fit model prediction is probabilistic. The black dashed line is the time-evolution of the mean. The three different grey shadings are the 1, 2, and $3\sigma$ intervals, respectively.}
\label{fig:lightcurve-with-best-fit-OU-model}
\end{figure}

\begin{figure}
\begin{center}
\includegraphics[width=8.0cm]{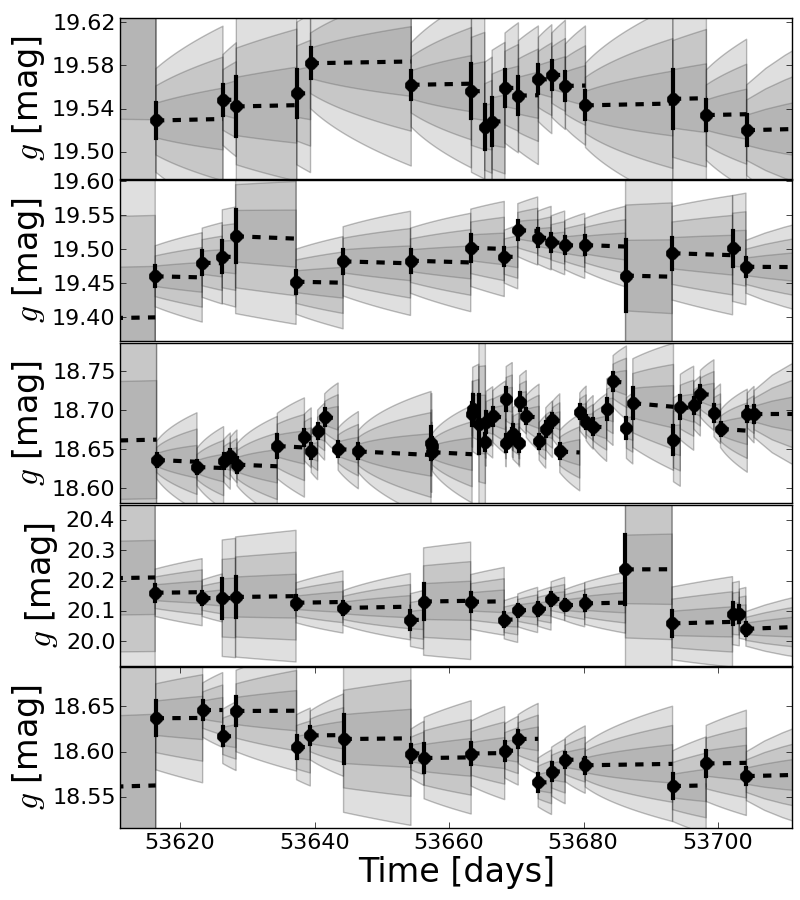}
\end{center}
\caption{Same as Fig.~\ref{fig:lightcurve-with-best-fit-OU-model} but now zoomed into the time interval from 53\,611 to 53\,711 days.}
\label{fig:lightcurve-with-best-fit-OU-model-zoom-in}
\end{figure}

Figure~\ref{fig:lightcurve-with-best-fit-OU-model} shows five randomly selected QSO lightcurves and their best-fit OU processes. We can nicely see the time evolution of mean and standard deviation of the probabilistic model prediction.\footnote{Mathematical details of the time evolution of mean and variance can be found in \citet{BailerJones2012}.} Note that the variance does not diverge with time because the OU process is stationary ($0<\phi_\textrm{OU}<1$).\footnote{A classical random walk is not stationary because $\phi=1$. In that case, the variance would indeed diverge.} Similarly, note that for high observed values, the mean evolves downwards, whereas for low observed  values the mean evolves upwards. This behaviour is called ``mean reversion'' and is also due to $0<\phi_\textrm{OU}<1$. We can see from Fig.~\ref{fig:lightcurve-with-best-fit-OU-model} that the best-fit OU processes provide good models over long time separations between observations. Figure~\ref{fig:lightcurve-with-best-fit-OU-model-zoom-in} shows the same but now resolving one of the time frames with many observations. Evidently, the best-fit OU processes are also excellent descriptions in that case. Note that the evolution of standard deviation of the model prediction does not start from 0 but from the error of the last observation.

\begin{figure}
\begin{center}
\includegraphics[width=8.0cm]{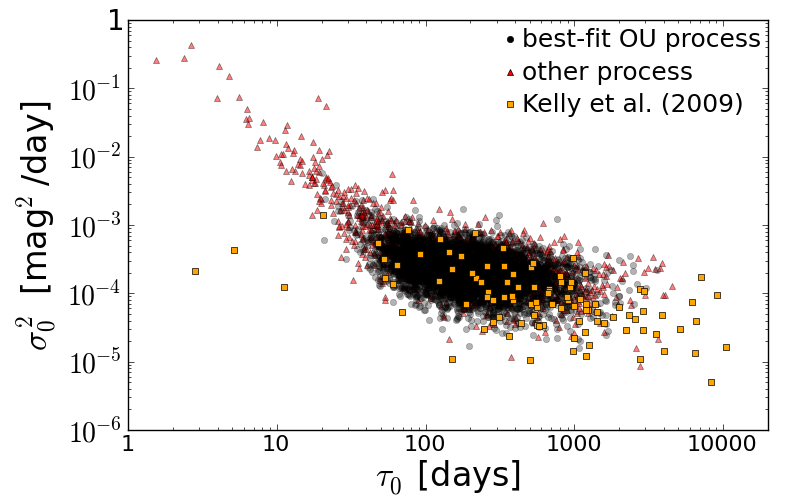}
\end{center}
\caption{Maximum-likelihood rest-frame parameters $\tau_0$ and $\sigma_0^2$ of the OU~process for all 6\,304~Stripe~82 QSO lightcurves. Black dots represent QSO lightcurves that are best described by an OU process, whereas red triangles represent those where the OU process is \textit{not} the best description. Orange squares represent 100~MACHO QSO lightcurves from \citet{Kelly2009}.}
\label{fig:best-fit-OU-parameters}
\end{figure}

\begin{figure}
\begin{center}
\includegraphics[width=8.0cm]{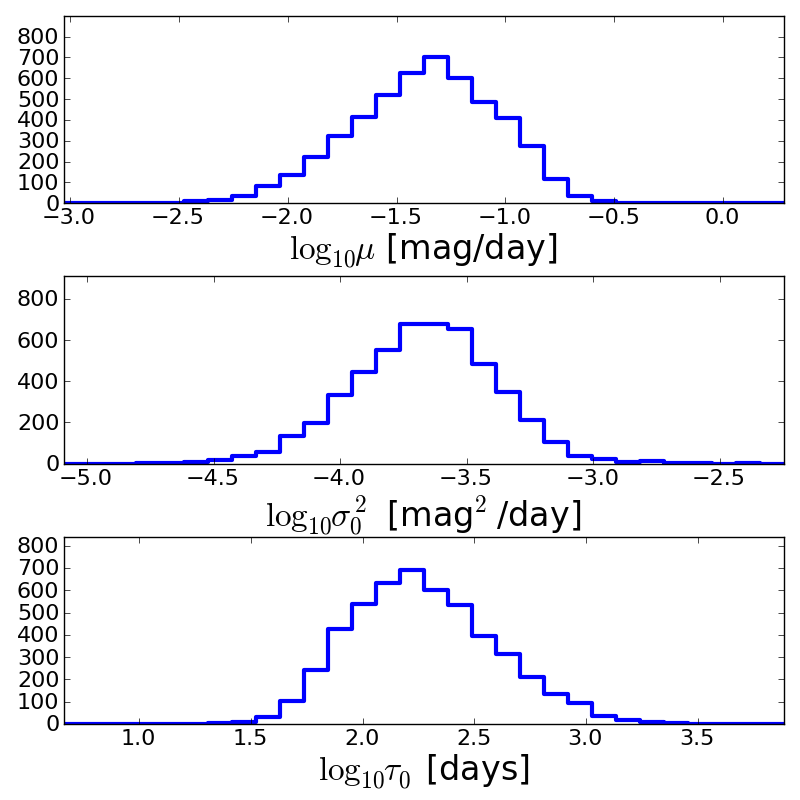}
\end{center}
\caption{Distribution of maximum-likelihood rest-frame parameters $\mu$, $\sigma_0^2$ and $\tau_0$ of the OU~process for 5\,047~QSO lightcurves that are best described by an OU process.}
\label{fig:best-fit-OU-parameters-hist}
\end{figure}

Let us now repeat the analysis of OU-process parameters that \citet{Kelly2009} conducted for 100 MACHO lightcurves. Our data sample is considerably larger and independent, such that this is also an independent test. Figure~\ref{fig:best-fit-OU-parameters} shows the distribution of best-fit rest-frame parameters
\begin{equation}
\sigma_0^2 = \sigma^2(1+z)
\end{equation}
and
\begin{equation}
\tau_0 = \tau/(1+z)
\end{equation}
for the OU process for a QSO lightcurve at redshift $z$. The distributions of $\tau_0$ and $\sigma_0^2$ for our 6\,304~QSO lightcurves appear to be consistent with the distribution of the same parameters for the 100 MACHO QSO~lightcurves as inferred by \citet{Kelly2009}. We too find the majority of relaxation timescales between 100 and 1000~days, while values of $\sigma_0$ are of the order of 0.01 as well. We note that there is a peculiar tail in parameter space, where QSO lightcurves have low relaxation timescales and large variation. These objects are lightcurves that prefer not being fit by the OU process. Figure~\ref{fig:best-fit-OU-parameters-hist} shows the distributions of best-fit parameters for the OU process. In detail, we obtain,
\begin{equation}
\log_{10}\mu = -1.37 \pm 0.34 \,\textrm{,}
\end{equation}
\begin{equation}
\log_{10}\sigma_0 = -1.83 \pm 0.15 \,\textrm{,}
\end{equation}
\begin{equation}
\log_{10}\tau_0 = 2.27\pm 0.33 \,\textrm{.}
\end{equation}
Note that the agreement between our results and those obtained by \citet{Kelly2009} provide an important consistency check, as our method and data are completely different and we did \textit{not} use results from \citet{Kelly2009} as prior distributions.

The typical values of relaxation timescales, $\tau$, of a few hundred days could be related to the typical estimated size of QSO accretion disks, which are a few hundred lightdays. As larger discs are brighter, this suggests a possible correlation of $\tau$ and brightness. However, we do not see noteworthy evidence of such a correlation in our data.

\subsection{Composite models}

In another attempt to construct better models of QSO lightcurves than the OU process, we combine a deterministic model with a stochastic process. In our opinion, the most reasonable choice is to combine a sinusoidal model with a stochastic process. When combining the sinusoid with a stochastic process, we can drop the offset parameter of the sinusoid and let the stochastic process take care of it. Furthermore, we no longer employ the periodogram prior on the sinusoid's period, since, first, the model has now two components generating variability, and, second, we no longer aim for conservative evidence against this model. Instead, we adopt a prior distribution that is flat in $P$, ranging from periods of 1 day to 10\,000 days.

\begin{table*}
\begin{centering}
\begin{tabular}{lcrrrrr}
\hline
model & 
\rotatebox{80}{number of model parameters} & 
\rotatebox{80}{no decisive evidence against it at $3\sigma$} &
\rotatebox{80}{no decisive evidence against it at $5\sigma$} & 
\rotatebox{80}{no decisive evidence against it at $7\sigma$} & 
\rotatebox{80}{no decisive evidence against it at $10\sigma$} & 
\rotatebox{80}{max evidence}\\
\hline\hline
constant                                                  & 1 & 0 & 0 & 0 & 0 & 0 \\
constant plus noise                               & 2 & 58 & 63 & 73 & 97 & 3 \\
sinusoid (flat prior in $P$)                              & 4 & 30 & 36 & 47 & 62 & 2 \\
sinusoid (flat prior in $\log P$)                         & 4 & 20 & 23 & 28 & 39 & 2 \\
sinusoid (periodogram prior)                              & 4 & 26 & 31 & 38 & 50 & 4 \\
two sinusoids (flat priors in $P$)                        & 7 & 33 & 37 & 48 & 62 & 8 \\
sinusoid (flat prior in $P$) plus noise                  & 5 & 467 & 584 & 722 & 940 & 146 \\
\hline
Wiener process (random walk)                                & 2 & 2251 & 2869 & 3438 & 4225 & 6 \\
OU process (damped random walk)                               & 3 & 6033 & 6082 & 6109 & 6139 & 3462 \\
Gaussian CAR(2) process ($\phi_1=\phi_2=0.1$)             & 2 & 0 & 0 & 0 & 1 & 0 \\
Gaussian CAR(2) process ($\tau_1=\tau_2$)                 & 3 & 1 & 2 & 2 & 3 & 0 \\
Gaussian CAR(2) process ($\tau_1\neq\tau_2$)              & 4 & 5 & 5 & 5 & 5 & 5 \\
Cauchy CAR(1) process ($\phi=1$)                          & 2 & 77 & 102 & 154 & 336 & 36 \\
symmetric stable CAR(1) process ($\phi=1$, $\beta=0$)     & 3 & 0 & 0 & 0 & 0 & 0 \\
stable CAR(1) process ($\phi=1$)                          & 4 & 0 & 0 & 0 & 0 & 0 \\
Gaussian CARMA(1,1) process                       & 4 & 0 & 0 & 0 & 0 & 0 \\
constant + Gaussian ARCH(1) process                     & 3 & 236 & 282 & 321 & 379 & 64 \\
sinusoid + Gaussian ARCH(1) process                     & 6 & 902 & 1212 & 1549 & 2138 & 155 \\
Gaussian CAR(1)-ARCH(1) process                     & 4 & 2005 & 2427 & 2884 & 3492 & 199 \\
Gaussian CAR(1)-GARCH(1,1) process                     & 5 & 3669 & 4140 & 4527 & 4948 & 450 \\
Gaussian CARMA(1,1)-GARCH(1,1) process           & 6 & 1575 & 2131 & 2799 & 3723 & 49 \\
\hline
sinusoid (flat prior in $P$) + Wiener process    & 5 & 11 & 12 & 12 & 12 & 7 \\
sinusoid (flat prior in $P$) + OU process    & 6 & 6028 & 6070 & 6105 & 6144 & 1706 \\
\hline
\end{tabular}
\end{centering}
\caption{Same as in Table~\ref{table:results-model-comparison} but now including models that combine a sinusoid with a stochastic process.}
\label{table:results-composite-models}
\end{table*}

Table~\ref{table:results-composite-models} shows how the model assessment changes, if we take such composite models into consideration. There are two interesting conclusions: First, adding a sinusoid to the Wiener process decreases the model performance drastically. Apparently, adding the sinusoid component does not improve the description of QSO lightcurves and the additional model parameters are penalised accordingly. Second, adding a sinusoidal component to the OU process creates a composite model that is highly competitive to the OU process alone, despite increasing the number of model parameters from 3 to 6. For 6\,028 QSO lightcurves, no other model can provide decisive evidence against this model at any relevant confidence level, while even 1\,706~of all lightcurves (27\%) are best described by this model. Nevertheless, twice as many QSO lightcurves are still best described by an OU process without sinusoidal modulation.

\begin{figure}
\begin{center}
\includegraphics[width=8.0cm]{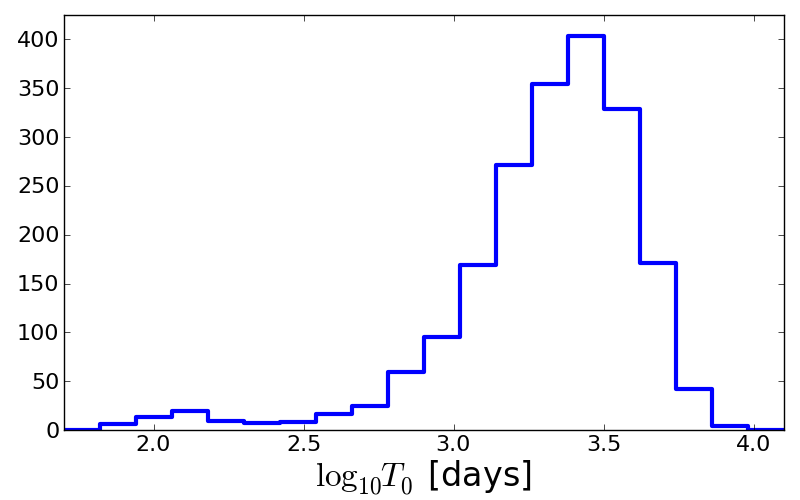}
\end{center}
\caption{Distribution of best-fit sinusoidal periods for those 1\,724 QSO lightcurves that are best described by a combination of sinusoid and OU process. We quote rest-frame periods that have been corrected for redshift.}
\label{fig:best-fit-period-sinus-OU-combi}
\end{figure}

Figure~\ref{fig:best-fit-period-sinus-OU-combi} shows the distribution of best-fit periods of QSO lightcurves favouring a combination of sinusoid and OU process. Typically, periods appear to be of the order of 1\,000 days, while approximately 90\% of periods are between 500 and 5\,000 days. This corresponds to long-term variations over several years and additional data are necessary to investigate this in more detail. For instance, continued observations of these targets by PanSTARRS may confirm the long-term variation of these sources.

\begin{figure}
\begin{center}
\includegraphics[width=8.0cm]{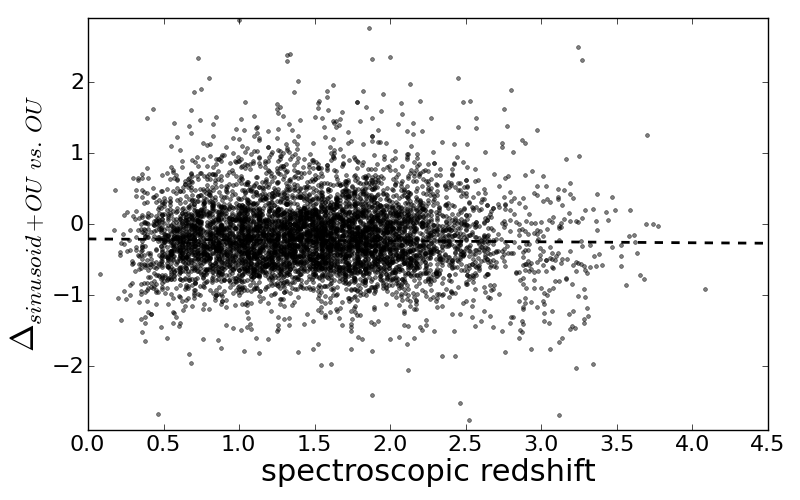}
\end{center}
\caption{Preference of OU process plus sinusoid over OU process alone as a function of spectroscopic redshift. The best-fit linear trend has a negative slope but it is only of $2.4\sigma$ confidence.}
\label{fig:trend-Delta-vs-redshift}
\end{figure}

\begin{figure}
\begin{center}
\includegraphics[width=8.0cm]{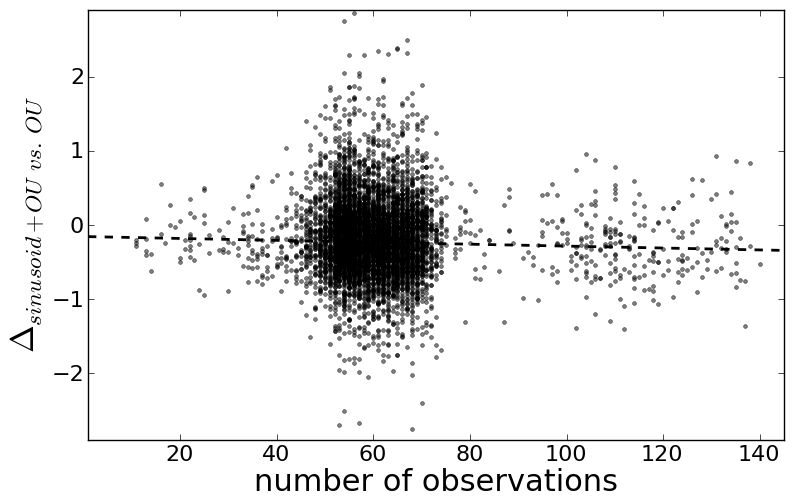}
\end{center}
\caption{Preference of OU process plus sinusoid over OU process alone as a function of number of observations. The best-fit linear trend has a negative slope with $5.2\sigma$ confidence.}
\label{fig:trend-with-Nobs}
\end{figure}

Let us test whether the preference for an additional sinusoidal component correlates with the available astrophysical information about these QSOs. Figure~\ref{fig:trend-Delta-vs-redshift} shows that there is no indication that preference for an additional sinusoid depends on spectroscopic redshift. Figure~\ref{fig:trend-with-Nobs} shows that there is a weak dependence on the number of observations. QSO lightcurves with more observations slightly tend to favour the OU process alone, without an additional sinusoid. This trend is very weak but could indicate that we do not have enough data to rule out the composite model.

Testing the preference for an additional sinusoid for a dependence on the QSO's brightness or colour is complicated by the large range of redshifts of the QSOs (c.f.~Fig.~\ref{fig:sample-distribution-z-Nobs}) and the required $K$ correction. Due to this large redshift range, absolute magnitudes computed from the apparaent $g$ would correspond to very different rest-frame wavelengths and would thus not be comparable. A similar argument applies to colours. However, we do not want to make a $K$ correction because we do not want to make any assumption on QSO spectra. Instead, we bin the data into narrow redshift slices of width $\Delta z=0.025$. QSOs with $z<0.3$ or $z>3.3$ are not considered here, as they are too sparse to allow a populated slicing. We then fit a linear relation to all redshift bins, where each redshift bin may have its own offset parameter accounting for the appropriate $K$ correction at that redshift, but all bins are simultaneously fit by an identical slope parameter, $b$. For the $g$-band magnitude, we obtain a slope of
\begin{equation}
b_g = -0.1420 \pm 0.0061 \,\textrm{,}
\end{equation}
and for $g-r$ colour, we obtain a slope of
\begin{equation}
b_{g-r} = -0.070 \pm 0.033 \,\textrm{.}
\end{equation}
These numbers provide circumstantial evidence that QSOs lightcurves which prefer an OU process with an additional sinusoid tend to be brighter and bluer than other lightcurves which prefer the OU process alone. Given the fact that bluer QSOs tend to be brighter due to the thermal origin of their radiation \citep[e.g.][]{Giveon1999,Trevese2001,Geha2003,Schmidt2012}, this could be interpreted as an indication that the preference of periodic oscillations may correlate with an increase in ``temperature'' of the emission region.

\section{Conclusions and discussion}
\label{sect:conclusions}

Given the results of Table~\ref{table:results-model-comparison}, we conclude that the overwhelming majority of QSO lightcurves are certainly not just due to Gaussian variations about a constant magnitude, in agreement with results of \citet{Sesar2007}. Furthermore, the overwhelming majority of QSO lightcurves do not exhibit any indication of sinusoidal variability. Indeed, our results (Table~\ref{table:stochasticity-assessment}) suggest that QSO lightcurves are not well described by simple deterministic models but instead are of stochastic nature, since we only find 29~out of 6\,304~QSO lightcurves where there is decisive evidence against all stochastic models. Furthermore, a substantial fraction of QSO lightcurves appear to exhibit stochasticity not only on the time-evolution of the mean value but also on the time-evolution of the variance itself (cf.~evidence for models with ARCH or GARCH component in Tables~\ref{table:results-model-comparison} and~\ref{table:results-composite-models}).

What is causing the stochasticity? \citet{Kelly2009} provide an argument that this stochastic nature could arise through thermal fluctuations in the accretion disk of the quasar, namely magnetic turbulence. There could also be flares in the accretion disk, caused by inhomogeneous infall of matter. Last but not least, there is also the possibility that QSO variability is governed by complex, nonlinear processes, which are actually predictable but cannot be well described by our simple deterministic models. 

\citet{Kelly2009} found the OU process to be a good description of 100 QSO lightcurves from the MACHO data. This result has been confirmed for other data by other authors \citep[e.g.][]{Kozlowski2010,MacLeod2010}. Furthermore, \citet{Kelly2009} speculated whether extensions of the OU process could lead to better models. We have tested both these hypotheses in the Bayesian framework. We have considered several possible extensions of the OU process and several alternative models. However, there is decisive evidence against almost all of these extensions for the majority of QSO lightcurves. We find only one extension that is competitive with the OU process, and that is the OU process combined with a sinusoidal oscillation. 1\,706~out of 6\,304~QSO lightcurves favoured this model, while most of the rest favoured the simple OU process. These QSOs seem to be brighter and bluer, which may indicate a relation with increasing temperature and thermal activity. Furthermore, for the vast majority of QSO lightcurves in our sample there is no evidence that multiple OU processes \citep{Kelly2011} provide better descriptions than a single OU process (Table~\ref{table:multiple-OU}). Comparing the best-fit OU parameters with results of \citet{Kelly2009}, we find consistent results obtained from independent methods and different data, which provides an important sanity check.

Possible reasons why the OU process appears to be such a good description of QSO lightcurves have already been offered in the literature:
\begin{enumerate}
\item The OU process approximates the shape of the observed power spectrum of QSO lightcurves \citep{Kelly2009,Kelly2011}.
\item The OU process is a solution to the linear stochastic diffusion equation, which may account for the astrophysical processes taking place in the QSO accretion disk \citep{Kelly2011}.
\end{enumerate}
Moreover, our results enable us to support the OU process by a plausibility argument: Given the finite energy supply of QSOs, a stochastic process must be stationary in order to be astrophysically reasonable.\footnote{With ``stationary'' we mean that the variance should not diverge with time. Although a finite segment of a lightcurve could well appear to be non-stationary.} We can also argue that it should be a Gaussian process, if the QSO emission is a linear superposition of numerous ``small'' emission events, e.g., in the QSO accretion disk, which have finite variance (due to finite energy supply) and are statistically independent.\footnote{These three assumptions -- linear superposition, finite variance, and statistical independence -- enable us to invoke the Central Limit Theorem (see also Appendix~\ref{appendix:CLT}).} Occam's razor then calls for the simplest model that is Gaussian and stationary. This simplest model is the Gaussian CAR(0) process -- a deterministic mean plus additional noise. We considered two examples, namely constant plus noise and sinusoid plus noise, and neither model described the lightcurves well. The next simplest Gaussian and stationary process is the CAR(1) process -- the OU process. Consequently, the OU process is the simplest model that is Gaussian and stationary and fits the data well.

We of course cannot claim that QSO lightcurves are generated by OU processes. There may be more complicated extensions of the OU process or entirely different (deterministic or stochastic) models that could provide even better descriptions. Moreover, QSO variability data probing different domains, e.g., in brightness, redshift, or time sampling, may lead to different results. In fact, results of \citet{Zu2013} suggest that there may be deviations from the OU process on very short timescales. However, we may conclude that from those models that we have considered here the OU process is by far the best model for QSO lightcurves from SDSS Stripe~82.

\paragraph{Acknowledgements.} We thank Gordon Richards, J\"org-Uwe Pott, Nina Hernitschek, Knud Jahnke, and Christian Fendt for discussions and helpful comments on the interpretation of our results.

\bibliographystyle{aa}

\def\physrep{Phys. Rep.}%
\def\apjs{ApJS}%
\def\apj{ApJ}%
\def\aj{AJ}%
\def\aap{A\&A}%
\def\aaps{A\&AS}%
\def\mnras{MNRAS}%
\def\araa{ARA\&A}%
\bibliography{bibliography}

\appendix

\section{Stable distributions in a nutshell}
\label{app:stable-PDFs}

This section provides a brief introduction to the familiy of stable distributions. These distributions play an important role, e.g., in the field of financial math.

\subsection{Generalised central-limit theorem}
\label{appendix:CLT}

Considering a real-valued random variate $Y$, which is the sum of $N$ real-valued random variates $X_n$,
\begin{equation}
Y = X_1 + X_2 + \ldots + X_N \,\textrm{,}
\end{equation}
the central-limit theorem states that the probability density function of $Y$, $p(y)$, converges to a Gaussian in the limit of $N\rightarrow\infty$, if and only if two conditions are satisfied:
\begin{enumerate}
\item All random variates $X_n$ are statistically independent of each other.
\item All random variates $X_n$ are drawn from distributions $p_n(x)$ which have finite mean and variance.
\end{enumerate}
The central-limit theorem is one of the primary reasons why the Gaussian distribution is so popular -- the other reason being the Gaussian's mathematical simplicity.

\citet{Gnedenko1954} investigated what happens if we drop the second condition of finite mean and variance of the $X_n$. They showed that in this case, $p(y)$ no longer approaches a Gaussian but rather a stable distribution. This is called the ``generalised central-limit theorem''. If one accepts the ``normal'' central-limit theorem as an argument in favour of the Gaussian distribution, one also has to accept the generalised central-limit theorem as an argument in favour of stable distributions. Which form of the central-limit theorem applies in practice, depends on the specific situation under consideration.

\subsection{Properties of stable distributions}

\begin{figure}
\includegraphics[width=8.0cm]{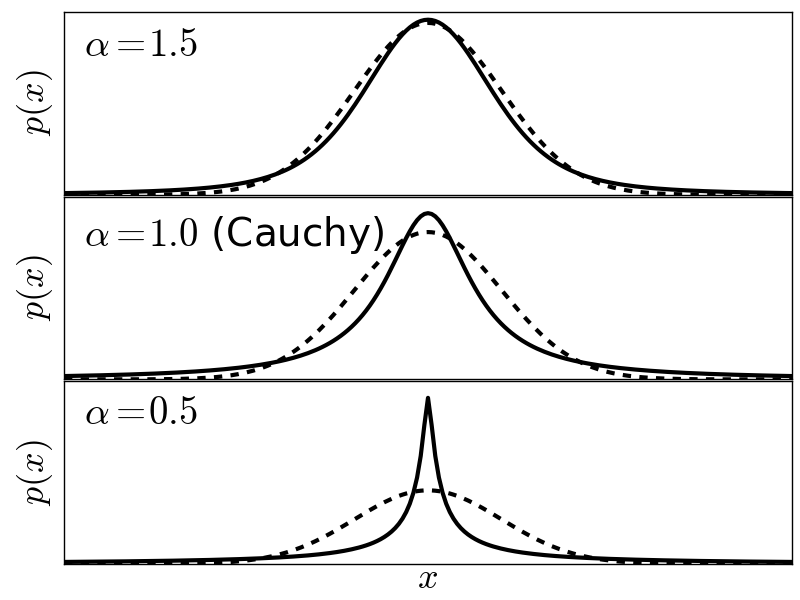}
\caption{Examples of stable probability density functions that have been evaluated through numerical integration. In all cases, $\delta$ and $\gamma$ are identical, $\beta=0$, and the dashed curve always shows a Gaussian for comparison's sake. Solid lines from top to bottom have values of $\alpha$ that are 1.5, 1, and 0.5. As $\alpha$ decreases, the tails become heavier and the peak becomes sharper when compared to the Gaussian.}
\label{fig:show-stable-PDFs}
\end{figure}

Here we give a brief overview over the most important properties of stable distributions. This overview is by no means exhaustive or complete, we only mention those properties that are relevant to this article.
\begin{enumerate}
\item Parametrisation: Stable distributions are fully described by four quantities:
\begin{itemize}
\item $\alpha\in(0,2]$, regulating the heavyness of the tails and the sharpness of the peak,
\item $\beta\in[-1,1]$, regulating the asymmetry ($\beta=0$ is symmetric),
\item $\gamma\in(0,\infty)$, regulating the width of the distribution,
\item $\delta\in(-\infty,\infty)$, regulating the location (if $\beta=0$, $\delta$ is the location of the peak).
\end{itemize}
\item Infinite moments: As soon as $\alpha <2$, the second moment (variance) is infinite. As soon as $\alpha < 1$, also the first moment (mean) is infinite. Only the zeroth moment (normalisation) is finite for all stable distributions.\footnote{At first glance, the notion of infinite mean/variance may appear odd. Evidently, any finite data sample must have finite sample-mean and sample-variance. On second thought, however, one should realise that distributions with infinite mean or variance are quite common in astronomy (e.g., power-law distributions, Schechter functions).}
\item Characteristic function: In order to fit a distribution to observed data, one needs to evaluate the probability density function (PDF) in order to compute the data's likelihood. Unfortunately, stable distributions usually do not have an analytic PDF. Only the characteristic function $\varphi(t)$ is given analytically,
\begin{equation}
\varphi(t|\alpha,\beta,\gamma,\delta) = 
\exp\left[it\delta-|\gamma t|^\alpha\,(1-i \beta\,\textrm{sgn}(t)\Phi)\right] \,\textrm{,}
\end{equation}
where
\begin{equation}
\Phi = \left\{\begin{array}{ccc}
\tan(\pi\alpha/2) & \Leftrightarrow & \alpha\neq 1 \\
-(2/\pi)\log|t| & & \textrm{else}
\end{array}\right.
\end{equation}
from which the PDF must be evaluated by Fourier transformation. Figure~\ref{fig:show-stable-PDFs} shows examples of stable PDFs. In practice, this makes likelihood evaluations computationally expensive. There are only four special cases, where the PDF exists analytically:
\begin{itemize}
\item Gaussian distribution ($\alpha=2$, $\beta=0$),
\item Cauchy distribution ($\alpha=1$, $\beta=0$),
\item L\'evy distribution ($\alpha=\frac{1}{2}$, $\beta=1$),
\item inverse L\'evy distribution.
\end{itemize}
\item Sampling: Although evaluating the PDF is numerically expensive, drawing random samples from a stable distribution is simple and has very low computational cost.
\item $\alpha$-stability: Let $X$ and $Y$ be two stable random variates with identical $\alpha$. Then, also the random variate $Z=X+Y$ is stable with identical $\alpha$ and the other parameters of the stable distribution of $Z$ are:
\begin{eqnarray}
\beta_Z = \frac{\beta_X\gamma_X^\alpha + \beta_Y\gamma_Y^\alpha}{\gamma_X^\alpha +\gamma_Y^\alpha} \label{eq:stability-beta} \\
\gamma_Z^\alpha = \gamma_X^\alpha +\gamma_Y^\alpha \\
\delta_Z = \delta_X + \delta_Y
\end{eqnarray}
The Gaussian distribution ($\alpha=2$, $\beta=0$) is a well-known example.
\end{enumerate}

\end{document}